\documentstyle[preprint,aps]{revtex}
\def\beq{\begin{equation}}
\def\eeq{\end{equation}}
\tighten
\begin{document}

\title
{\bf Continuum HFB calculations with finite range pairing 
  interactions}

\author
{M. Grasso,$^1$ N. Van Giai,$^1$ N. Sandulescu,$^{2, 3}$}

\address
{$^1$ Institut de Physique Nucl\'eaire, IN2P3-CNRS,
Universit\'e Paris-Sud, 91406 Orsay Cedex, France\\
$^2$ Institute for Physics and Nuclear Engineering,
 P.O. Box MG-6, 76900 Bucharest, Romania\\
$^3$ Research Center For Nuclear Physics, Osaka University, Ibaraki, Osaka,
567-0047, Japan}
 
\maketitle

\begin{abstract}
A new method of calculating pairing correlations in coordinate space with
finite range interactions is presented. In the Hartree-Fock-Bogoliubov (HFB)
approach the mean field part is derived from a Skyrme-type force whereas the
pairing field is constructed with a Gogny force. An iterative scheme is used
for solving the integro-differential HFB equations via the introduction of a
local equivalent potential. The method is illustrated on the case of the
nucleus $^{18}$C. It is shown that the results are insensitive to the cut
off value in the quasiparticle spectrum if this value is above 100 MeV. 
%Some Hartree-Fock-Bogoliubov (HFB) calculations for the neutron-rich isotope
%$^{18}$C are presented. The aim of the letter is to study the convergence of
%the results with respect to the cutoff which is chosen to delimitate the
%continuum phase space included in the HFB calculations. The mean field is
%constructed with a Skyrme interaction. The pairing field is calculated with
%both a zero-range and a finite range interactions. 
%It is shown that while
%the use of a zero-range interaction in the pairing channel causes the
%divergences of the results with respect to the increase of the cutoff the
%use of the finite range Gogny interaction furnishes a natural quenching at
%high energy. The convergence of the results is achieved around the cutoff
%value of 100 MeV for the HFB quasiparticle continuum spectrum. 
\end{abstract}

\vspace{1cm}

The treatment of pairing correlations is very important for the 
description of the properties of weakly bound nuclei situated close to the
drip lines. The Hartree-Fock-Bogoliubov (HFB) method \cite{rischu} is the 
the commonly adopted approach for treating 
 self-consistently both the mean field
contributions and the pairing correlations. Various effective 
%phenomenological
 interactions are
available and can be used in the mean field and 
pairing channels of the HFB equations. They are of two types, namely 
%They are generally chosen of
zero-range forces \cite{sk1,vab,bef,giai,lyon} and finite range forces
whose typical representatives are the Gogny forces \cite{gog}. 

Zero-range forces are widely used because  
the self-consistent equations can be conveniently
solved in coordinate space. In loosely bound systems such as unstable
nuclei, continuum effects in the pairing channel become important since 
their Fermi energies are typically close
to zero and the pairing correlations can easily populate the continuum.
Then, working in coordinate space is an advantage 
because the continuum effects can be accurately treated \cite{doflo,grali}.
On the other hand, a zero-range pairing interaction has the well-known
pathology of producing diverging contributions if no cut off is imposed on
the quasiparticle space. This cut off must be an inherent part of the
phenomenological zero-range interaction \cite{dona}, but it is not clear
which cut off value must be adopted for a given interaction and a given
nucleus. We note that some regularization scheme has been proposed recently
\cite{bul1} for dealing with the question of HFB equations with
zero-range pairing interactions.  

Finite range pairing interactions of course do not require in principle a
truncation of the quasiparticle space, even though in practical calculations
the summations are restricted to quasiparticle energies below some cut off
energy $E_{c.o.}$. If $E_{c.o.}$ is large enough the results no longer
depend on its precise value, in contrast to the case of zero-range
interactions. It is desirable therefore to have a method which combines the
advantages of solving HFB equations in coordinate space and of describing
pairing correlations with a finite range interaction. It is the purpose of
this paper to propose such a method and to illustrate it by comparing
results obtained with Skyrme and Gogny interactions.

We treat the continuum states exactly as explained in Ref. \cite{grali}
where Skyrme-type interactions were used both in the mean field and pairing
channels. Here, we keep a Skyrme force for the Hartree-Fock (HF) mean field
but the pairing interaction is taken as a Gogny force. 
%We show that the use of a Gogny interaction in the pairing channel
%eliminates the problem of the arbitrary choice of a cutoff.
% We adopt the local equivalent potential method \cite{vau} 
%to reduce the integro-differential HFB equations into
%differential ones by calculating an equivalent local potential from the non
%local Gogny potential. 
The calculations are done for the neutron-rich
nucleus $^{18}$C with the Skyrme interaction SLy4 \cite{lyon} in the mean field
channel and the Gogny interaction D1S \cite{ro} in the pairing channel.

We would like to stress here that with these calculations we do not want to
make any predictions on the properties of the chosen nucleus: to do this a
fit of the parameters of the Skyrme force to be used with the Gogny pairing
force must be done.
However, the purpose of this work is not to make
predictions on the properties of the nucleus $^{18}$C but to present and
show the advantages of a method of solving HFB equations where a zero-range
and a finite range interactions are combined together. 
%Two are the purposes of this work. On one side we want to introduce the
%local equivalent potential method that was already used for the solution of
%non local Schroedinger equations in \cite{vau} to solve the coupled
%integro-differential HFB equations. On the other side we want to show that
%the use of a finite range interaction in the pairing channel of these
%equations eliminates the divergence of the pairing correlations with the
%increase of the cutoff energy. 

For clarity we recall the general form of the coupled 
integro-differential HFB equations in
coordinate representation \cite{doflo}:
%is the following:
\beq
\begin{array}{c}
\int d^3 {\mathbf r'} \sum_{\sigma'} \left( \begin{array}{cc} 
h( {\mathbf r} \sigma , {\mathbf r'} \sigma' ) & 
\tilde{h}( {\mathbf r} \sigma , {\mathbf r'} \sigma' \\
\tilde{h}( {\mathbf r} \sigma , {\mathbf r'} \sigma' ) & 
-h ( {\mathbf r} \sigma , {\mathbf r'} \sigma' \end{array} \right)
\left( \begin{array}{c} \phi_1 (E, {\mathbf r'} \sigma' ) \\
\phi_2 (E, {\mathbf r'} \sigma' )\end{array} \right) = \\

   \\
\left( \begin{array}{cc} E+E_F & 0 \\ 0 & E-E_F \end{array} \right) 
\left( \begin{array}{c} \phi_1 (E, {\mathbf r} \sigma ) \\
\phi_2 (E, {\mathbf r} \sigma )\end{array} \right)~,
\end{array}
\label{1st}
\eeq
where $E_F$ is the Fermi energy, $\phi_1$ and $\phi_2$ are the upper and
lower components of the HFB quasiparticle wave functions, respectively. In
Eq.(\ref{1st}) the HF operator is a sum of kinetic and mean field components:
\beq
h({\mathbf{r}} \sigma , {\mathbf{r'}} \sigma' )= T(
{\mathbf{r}} , {\mathbf{r'}}) \delta_{\sigma  \sigma' } +
\Gamma ({\mathbf{r}} \sigma, {\mathbf{r'}} \sigma' )~,
\label{2st}
\eeq
\beq
\Gamma({\mathbf{r}} \sigma, {\mathbf{r'}} \sigma' ) =
\int d^3  {\mathbf{r_1}} d^3 {\mathbf{r _2}} \sum_{\sigma_1 \sigma  _2}
V({\mathbf{r}} \sigma , {\mathbf{r_1}} \sigma _1 ;
{\mathbf{r'}} \sigma'  , {\mathbf{r _2}} \sigma  _2 )
\rho({\mathbf{r _2}} \sigma  _2 , {\mathbf{r _1}} \sigma _1 )~.
\label{3st}
\eeq
When the HF interaction $V$ is a Skyrme force, the mean field $\Gamma$ is
just a functional of the local particle density $\rho({\bf r})$ and of
its derivatives. Then, the HF operator $h$ is a differential operator, which
is a familiar property of Skyrme-HF models. On the other hand, the pairing
interaction $V_{pair}$ is a local force of finite range, and the pairing
operator $\tilde{h}({\mathbf{r}} \sigma, {\mathbf{r'}} \sigma' )$ remains a
fully non-local kernel obtained by folding $V_{pair}$ with the pairing
density $\tilde{\rho}$:
\beq
\tilde{h}({\mathbf{r}} \sigma, {\mathbf{r'}} \sigma' ) =
\int d^3 {\mathbf{r_1}} d^3 {\mathbf{r _2}} \sum_{\sigma _1 \sigma  _2}
2 \sigma'  \sigma'  _2 V_{pair}({\mathbf{r}} \sigma , {\mathbf{r'}} -
\sigma'  ;
{\mathbf{r _1}} \sigma  _1 , {\mathbf{r _2}} -\sigma  _2 )
\tilde{\rho}({\mathbf{r _1}} \sigma  _1 , {\mathbf{r _2}} \sigma _2 )~.
\label{4st}
\eeq
This is the new feature as compared to the equations discussed and solved in
Refs.\cite{doflo,grali}. 
%In Eq. (\ref{2st}) $T$ represents the kinetic contribution and $\Gamma$ the
%mean field. The latter is given by Eq. (\ref{3st}) where $V$ is the interaction
%chosen to construct the mean field and $\rho$ is the particle density.
%The pairing field $\tilde{h}$ is given by Eq. (\ref{4st}) where $V_{pair}$
%is the interaction chosen to construct the pairing field and $\tilde{\rho}$
%is the pairing density. 

%If one works 
We choose to work 
with positive energy quasiparticle states, and the particle and
pairing densities  are expressed as:
\beq
\rho({\mathbf r} \sigma, {\mathbf r'} \sigma )
= \sum_{0 < E_n < -E_F } \phi _2 (E_n, {\mathbf r} \sigma) \phi _2 ^* 
(E_n,{\mathbf r'} \sigma)+
\int_{-E_F}^{E_{c.o.}} dE \phi _2 (E, {\mathbf r} \sigma) 
\phi _2 ^* (E, {\mathbf r'} \sigma)~,
\label{13}
\eeq
\beq
\tilde{\rho} ({\mathbf r} \sigma, {\mathbf r'} \sigma )
= - \sum_{0 < E_n < -E_F } \phi _2 (E_n, {\mathbf r} \sigma) 
\phi _1 ^* (E_n,{\mathbf  r'} \sigma)-
\int_{-E_F}^{E_{c.o.}} dE \phi _2 (E, {\mathbf r} \sigma) 
\phi _1 ^* (E, {\mathbf r'} \sigma )~.
\label{14}
\eeq
In Eqs. (\ref{13}) and (\ref{14}) the
summations are over the discrete states that are situated in the region
 of the spectrum $0<E_n<-E_F$ whereas the integrals run over the
continuum part of the spectrum up to a chosen cut off. 

%With the use of zero-range interactions in both mean field and pairing
%channels Eqs. (\ref{1st}) become coupled differential equations. 
%Here we use a zero-range interaction in the mean field channel
%and a finite range interaction in the pairing channel, so that 
%Eqs. (\ref{1st}) remain integro-differential. 

We now take for $V_{pair}$ a Gogny force which contains a 
%Let us introduce the non local Gogny pairing field. 
%The Gogny interaction has a finite range part that is the 
sum of two
gaussians, a zero-range density-dependent part and a zero-range spin-orbit
part. 
%contribution that is formally the same as in the Skyrme interactions. 
Within the parametrisation D1S \cite{ro} that we
adopt in this work the zero-range density-dependent part does not
contribute. Let us explain the contribution of the gaussian terms. That of
the spin-orbit part is also included in the calculations presented below
although its effect is quite small. The finite range part of the interaction
is:  
%The finite range term reads as follows,
\beq
V_{pair}({\mathbf r_1} - {\mathbf r_2} )=\sum_{\alpha =1,2}
(W_{\alpha}+ B_{\alpha} {\mathbf P_{\sigma}} -
 H_{\alpha} {\mathbf P_{\tau}} - M_{\alpha} {\mathbf P_{\sigma}} {\mathbf
 P_{\tau}}) e^{-
{|{\mathbf r_1} -{\mathbf r_2} |^2} \over {\mu_{\alpha}^2}}~, 
\label{1}
\eeq
where $W_{\alpha}$, $B_{\alpha}$, $H_{\alpha}$, $M_{\alpha}$ and
$\mu _{\alpha}$ are parameters, ${\mathbf P_{\sigma}}$ and
${\mathbf P_{\tau}}$ are the spin and isospin exchange operators,
respectively. Then, the pairing operator (\ref{4st}) becomes \cite{dona}:   
%In ref. \cite{dona} it is shown 
%that by using the finite range part of the Gogny
%interaction the pairing field given by Eq. (\ref{4st}) 
%can be written in terms of the pairing density as follows,
\beq
\tilde{h}({\mathbf r} \sigma, {\mathbf r'} \sigma ) = \sum_{\alpha=1,2} e^{-
{|{\mathbf r}-{\mathbf r'}|^2} \over \mu_{\alpha}^2} \left[(W_{\alpha} -
H_{\alpha}) \tilde{\rho}({\mathbf r} \sigma, {\mathbf r'} \sigma )- 
(B_{\alpha} -
M_{\alpha}) \tilde{\rho}({\mathbf r'} \sigma, {\mathbf r} \sigma )\right]~.
\label{3}
\eeq
In the following we will need the multipole expansions of the gaussian form
factors:
\beq
e^{-
{|{\mathbf r}-{\mathbf r'}|^2} \over \mu_{\alpha}^2}= \sum_{LM} F_L ^{\alpha}
(r,r')
Y_{LM} (\hat{r}) Y_{LM}^* (\hat{r'})~.
\label{4}
\eeq

We restrict ourselves to the case of spherical symmetry, for simplicity.
Then, the general set of equations (\ref{1st}) can be solved for each
partial wave $(l,j)$ separately. We define the radial parts of the
quasiparticle wave functions $\Phi_i(E, {\bf r}\sigma)$ (i=1,2) by:
\beq
\Phi_i(E ljm, {\bf r}\sigma) = \frac{u_i(E lj, r)}{r} Y_{lm_l}({\hat r}) (l
m_l \frac{1}{2} \sigma \vert j m) \chi ( \sigma )~,
\label{wf}
\eeq
where $\chi(\sigma)$ is a spinor corresponding to a spin projection $\sigma$. 
%Let us expand in multipoles the gaussian term of the pairing field $\tilde{h}$,
%where $Y_{LM}$ represent the spherical harmonics. From Eq. (\ref{4}) it is easy
%to derive the expression for the functions $W$,
%\beq
%W_L ^{\alpha} (r,r')=2 \pi \int_0 ^{\pi} sin \theta d\theta P_L (cos \theta)
%e^{ -{ 1 \over \mu_{\alpha}^2} (r^2 +r'^2 -2 r r' cos \theta ) }~,
%\label{5}
%\eeq
%where $P_L$ are Legendre polynomials.
%
%Let us impose the spherical symmetry that allows us to separate the radial
%and the angular parts of the wave functions; 
%let us call $u_1$ and $u_2$ the radial
%parts of the upper and lower components of the HFB quasiparticle wave
%functions respectively 
%and let us write the pairing density $\tilde{\rho}$ as follows,
%\beq
%\tilde{\rho} ({\mathbf r} \sigma, {\mathbf r'} \sigma ) = - \sum_{nlj}
%\sum_{m_j m_l m'_l} { {u_2(nlj,r)} \over r } Y_{l m_l} (\hat{r}) (
%l m_l {1 \over 2} \sigma | j m_j ) { {u_1(nlj,r')} \over r' } Y_{l m'_l}^* 
%(\hat{r'}) (
%l m'_l {1 \over 2} \sigma | j m_j )~,
%\label{6}
%\eeq
%where $l$ and $j$ denote a given partial wave $(l,j)$ and $n$ counts the
%states in each $(l,j)$ channel; the summation over $n$ becomes an integration
%over the energy for the continuum states included in the calculation.
%By using Eqs. (\ref{4}) and (\ref{6}) and by applying the properties of the
%spherical harmonics it is straightforward to rewrite Eq. (\ref{3}) and
%derive the following expression
With the help of the definition (\ref{wf}) and the expansion (\ref{4}) it is
straightforward to obtain the multipole decomposition of the pairing field 
%for the pairing field  
$\tilde{h}({\mathbf r}
,{\mathbf r'})\equiv \sum_ {\sigma}\tilde{h}({\mathbf r} 
\sigma ,{\mathbf r'} \sigma)$:
\beq
\tilde{h}({\mathbf r} ,{\mathbf r'} )=\sum_{L_1 M_1} \tilde{h}_{L_1} (r,r')
Y_{L_1 M_1} (\hat{r}) Y^*_{L_1 M_1} (\hat{r'})~,
\label{8}
\eeq
%
%\beq
%\begin{array}{c}
%\tilde{h}({\mathbf r} ,{\mathbf r'} )=\sum_{\alpha =1,2} \sum_L W_L ^{\alpha}
%(r,r') { {2L+1} \over 4 \pi} \sum_{nlj} (2j+1) \\
%
%\\
%
%\left[ (H_{\alpha} -
%W_{\alpha}) { u_2(nlj,r) \over r} { u_1(nlj,r') \over r'} -
%(M_{\alpha} -
%B_{\alpha}) { u_2(nlj,r') \over r'} { u_1(nlj,r) \over r} \right] \\
%
%\\
%
%\sum_{L_1
%M_1} \left(
%\begin{array}{ccc} L & l & L_1 \\ 0 & 0 & 0 \end{array} \right)^2
%Y_{L_1 M_1}(\hat{r}) Y^*_{L_1 M_1}(\hat{r'})~.
%\end{array}
%\label{7}
%\eeq
%
%As it is possible to write
%one can extract the radial part of $\tilde{h}$ from Eq. (\ref{7}),
where 
\beq
\begin{array}{c}
\tilde{h}_{L_1}(r,r')=\sum_{\alpha=1,2} \sum_L F_L ^{\alpha}
(r,r') { {2L+1} \over 4 \pi} \sum_{nlj} (2j+1) \\

\\

\left[ (H_{\alpha} -
W_{\alpha}) { u_2(nlj,r) \over r} { u_1(nlj,r') \over r'} -
(M_{\alpha} -
B_{\alpha}) { u_2(nlj,r') \over r'} { u_1(nlj,r) \over r} \right] \\

\\

\left( \begin{array}{ccc} L & l & L_1 \\ 0 & 0 & 0 \end{array} \right)^2~.
\end{array}
\label{9}
\eeq
%The summations over $n$ in Eqs. (\ref{7}) and 
The summations over $n$ in Eq. (\ref{9}) become integrals
over the energy for the continuum states.

For each partial wave $(lj)$ one has to solve a system of two coupled
integro-differential equations whose general structure is:
\begin{eqnarray}
h u_1(r) + \int {\tilde h}(r, r') u_2(r') r'^2 dr' & = & (E + E_F)
u_1(r)~, \nonumber  \\
\int {\tilde h}(r, r') u_1(r') r'^2 dr' - h u_2(r) & = & (E - E_F) u_2(r)~.
\label{intdiff}
\end{eqnarray}
In the context of HF equations with finite range interactions it has been
shown by Vautherin and V\'en\'eroni \cite{vau} that one can transform the
HF integro-differential equation into a purely differential
equation  by introducing a so-called
trivially equivalent local potential. Here, we generalize this method to the
system of equations (\ref{intdiff}) by defining local equivalent potentials
$U_i(r)$ in the following way: 
\begin{eqnarray}
\int \tilde{h} (r,r') u_i(r') r'^2 dr' & = &  { 1 \over {u_i (r)}} (\int
\tilde{h} (r,r') u_i(r') r'^2 dr') u_i (r) \nonumber \\  \nonumber \\
& \equiv & U_i(r) u_i(r)  ~, \; i=1,2~.
\label{10}
\end{eqnarray}
Then, Eqs.(\ref{intdiff}) become formally a system of two coupled
differential equations where the potentials depend on the solutions and
therefore they must be solved iteratively. This is not a major problem since
the self-consistency requirement already leads to an iterative scheme.

%The pairing field $\tilde{h}$ given by Eq. (\ref{9}) is non local.
%We adopt the method of the local equivalent potential to
%calculate from it local potentials to be used in the
%HFB equations; they will be in this way differential equations and can be
%easily solved by using the Numerov algorithm.
%The local equivalent potential method
%has been introduced in \cite{vau} to solve
%non local Schroedinger equations.
%We refer to \cite{vau} for more details about the method.
%
%If $u_1$
%and $u_2$ are the radial parts of the upper and lower components of the HFB
%wave functions respectively
%we define a local equivalent potential for each of them as follows,
An additional difficulty comes from the fact that the local potentials
$U_i(r)$ have poles at
%The local potentials given by Eq. (\ref{10}) diverge in correspondence with
the nodes of the wave functions $u_i(r)$.
%For example, let us consider the resonant state corresponding to the bound
%occupied
%state $2s1/2$ in the Hartree-Fock (HF) limit for the isotope $^{18}$C.
%This state is situated in the
%continuum part of the HFB quasiparticle spectrum and it has an energy of about
%6.5 MeV (the energy of the resonant state is calculated by studying the
%behavior of the phase shift of the continuum wave functions as illustrated
%in \cite{grali}). The local equivalent potential is calculated for the wave
%function associated with the maximum of the occupation probability in the
%region of the
%resonance. Since the chosen state corresponds to the $2s1/2$ bound occupied
%state in the HF limit the lower component
%$u_2$ of its wave function has a node; it is situated at about 2.4 fm. We
%calculate the pairing field $\tilde{h}^L (r,r')$ for $L=0$ by using Eq.
%(\ref{9}); from it, a local equivalent potential for the wave function $u_2$
% is evaluated by using Eq. (\ref{10}). This potential is
%drawn in Fig. 1 and presents a divergence at about 2.4 fm.
In Ref. \cite{vau} a very simple and efficient method was proposed to
overcome this problem, based on the linearization of the local equivalent
potential around the poles.
%the divergences of the local potentials are
%eliminated in the following way:
Another equivalent potential
$U_{i} (\epsilon, r)$ is introduced; it is equal to $U_{i} (r)$ everywhere
except inside the intervals
$[r_0 - \epsilon , r_0 + \epsilon]$ where $r_0$ denotes a pole of
$U_i (r)$. Inside these intervals $U_i (\epsilon, r) $ is
chosen as a segment which joins the values of  $U_i(r_0 - \epsilon) $ and
$U_i(r_0 + \epsilon)$. The approximation is good if $\epsilon$ is small enough
not to wash out the shape of the potential. Thus, $U_i(r)$ 
is replaced by the following potential:
\beq
U_{i}^{ (n+1)} (\epsilon_{n+1} , r) = R_{\epsilon_{n+1}} 
\left\{
{1 \over {u _{i} ^{(n)} (r)}} \int r'^2 dr' \tilde{h} (r,r')
u _{i} (r')^{(n)} dr' \right\}~,
\label{13nn}
\eeq
where $R_{\epsilon_{n+1}}$ indicates the linear interpolation procedure 
described above and $n$ counts the iterations by which the HFB
equations are solved.
 The parameter $\epsilon$ depends on the iteration and
it is chosen so that $lim~ \epsilon _n =0$. 
At each iteration of the HFB scheme 
%self-consistent calculation 
we evaluate the
equivalent potentials of all quasiparticle states 
%for each energy, i.e. for each wave function  
by using the wave functions of the previous iteration
and we repeat this procedure until convergence. 

We compare now the HFB results for $^{18}$C obtained by using in the pairing
channel either the finite
range Gogny interaction D1S
%with those obtained with
or a zero-range interaction. In both cases the Skyrme interaction SLy4 is used
to construct the mean field. The quasiparticle continuum is fully treated
(no box boundary conditions) as described in Ref.\cite{grali}. 
As it was said before, the
problem with the use of a zero-range pairing interaction
is the divergence of the pairing correlations when one increases
the cut off energy $E_{c.o.}$. 
%which delimitates the continuum spectrum. 
This is illustrated 
in Fig.1 where the particle (top) and pairing (bottom)
densities of neutrons in $^{18}$C are plotted. The zero-range pairing
interaction in 
this case is\cite{dona}: 
%with the following zero-range density
%dependent interaction in the pairing channel  ,
\beq
V_{pair}({\mathbf r} , {\mathbf r'} ) = V_0 \delta ({\mathbf r} - {\mathbf r'} ) 
\left[
1- \left( { \rho({\mathbf r} )\over \rho_C } \right) ^{\gamma} \right]~,
\label{12}
\eeq
where $\rho({\mathbf r})$ is the particle density, and the
values of $V_0$ and
$\gamma$ are fixed from the parameters of SLy4 ($V_0$=-2488.9 MeV fm$^3$,
$\gamma$=1/6, with the choice $\rho_C$ = 0.133 fm$^{-3}$).
This particular choice of the parameters does not influence significantly
the convergence properties at high energies cut-offs discussed
here.

In Fig.1 different densities 
corresponding to different values of $E_{c.o.}$ are
plotted. 
%in correspondence with different cutoff values for the continuum phase space. 
We can observe that,  
while the particle density is almost
stable with respect to the cut off energy
the pairing density increases with $E_{c.o.}$ 
as it is expected. This indicates that the mean field properties are not
much affected by the enlarging of the continuum phase space 
%introduced in the calculations 
while the pairing properties are very sensitive to it. If
one moves the cut off towards higher values the pairing density continues
to increase and it never converges to a stable result. This problem is
eliminated when a finite range interaction is used to construct the pairing
field. 

We show in the upper part of Fig.2 the total energies 
of $^{18}$C calculated with the two
pairing interactions and for cut off values ranging from 70 MeV up to
120 MeV. We observe that the system is more and more bound when the cutoff
increases; this is due to the fact that pairing correlations become more and
more important.
While in the case of the zero-range interaction the total energy
continues to decrease in the chosen interval of cut off values, in the case
of the Gogny interaction the total energy converges and reaches a stable value
equal to -130.65 MeV at a cut off of 100 MeV. We can equivalently observe
this convergence of the energy by studying the trend of the pairing
correlation energies which are shown in the lower part of 
Fig. 2 for the same cut off values.
The pairing correlation energy is defined as follows:
\beq
E_P \equiv E(HF) - E(HFB)~,
\label{13n}
\eeq
where $E$ indicates the total energy of the nucleus. The quantity $E_P$
is the difference in the total binding energies between the HF
%mean field calculation
and HFB calculations; thus, it gives an estimation of the
amount of pairing correlations.
In the figure one can observe that
these energies always increase for the zero-range interaction indicating the
increasing of the amount of pairing correlations while they reach a stable
value equal to 16.04 MeV for
 the Gogny interaction. It is easy to understand why the results with the
latter interaction become stable for cut off energies around 100 MeV or
beyond. The shortest range of the two gaussian form factors is 0.7 fm, which
corresponds to 2.86 fm$^{-1}$ in momentum space, i.e., a kinetic energy of
about 160 MeV. The depth of the mean field potential being about 40 MeV one
can estimate that this kinetic energy corresponds to a quasiparticle energy
around 120 MeV.  

Another quantity that one can study as a function of the cut off energy is
the mean square radius of the pairing density. Indeed, the mean square
radius of the particle density must be stable against $E_{c.o.}$ because the
$u_2^2(r)$ functions entering Eq.(\ref{13}) are negligible for large
quasiparticle energies. In contrast, the $u_1(r)u_2(r)$ factors of
Eq.(\ref{14}) do not decrease so fast with increasing quasiparticle
energies. Let us define: 
%that o that we tested are the momenta $M_{\alpha}$ defined as follows,
\beq
\langle r^{\alpha} \rangle \equiv \int r^2 dr \tilde{\rho} (r) r^{\alpha}~.
\label{14n}
\eeq
In Fig. 3 we show $\langle r^{2} \rangle$ for
the two pairing interactions.
Again, while $\langle r^{2} \rangle$ always increases
in the case of the zero-range interaction it
reaches a stable value of about 34.5 fm$^2$ 
at a cut off of 95 MeV in
the  case of the Gogny interaction. This convergence indicates that the
pairing density does not change when the cut off is moved beyond the value of
95 MeV. 

In this work we have presented a new method for solving HFB equations in
coordinate space with finite range pairing interactions. This may be useful
in systems where the chemical potential is close to zero and an accurate
treatment of the quasiparticle continuum is required. As an illustration of
the method we have solved the HFB equations for a neutron-rich nucleus,
using the finite range force D1S as pairing interaction. Because our main
purpose here is to discuss the respective behaviour of zero-range and finite
range interactions in the pairing channel, we have kept in the present
application the Skyrme-type force SLy4 to generate the HF mean field. A
further step toward full self-consistency with the same finite range force
for calculating the mean field and the pairing field can be made, using the
same technique of local equivalent potential also for the mean field as it
was already done in the HF context \cite{vau}. Work in this direction is in
progress.

%In this letter we have discussed the problem of the sensibility of the HFB
%results with respect to the chosen energy cutoff when the continuum states
%are included in the calculations depending on which pairing interaction is
%used. 
%On one side we 
%have exploited the local equivalent potential method to calculate 
%equivalent potentials from the non local Gogny potential in order to reduce 
%the coupled integro-differential HFB 
%equations into coupled differential equations
%when the Gogny finite range interaction is used in the pairing channel.
%On the other side 
%we have shown that the obtained results reach the stability around a cutoff
%of 100 MeV when the Gogny interaction is used in the pairing channel, while
%they diverge if a zero-range interaction is chosen. 

\vspace{1cm}

The authors wish to thank P.F. Bortignon, P. Schuck and N. Vinh Mau 
for fruitful discussions. One of us (M.G.) 
%acknowledges that this work was supported by 
is a recipient of a European Community
Marie Curie Fellowship. Two of us (N.V.G. and N.S.) acknowledge that this
work was done in the framework of IN2P3(France)-IPNE(Romania) Collaboration.

%\newpage

%\begin{figure}
%\caption{The local equivalent potential for the wave function in Fig. 1
%calculated from the non local Gogny potential with the parametrisation D1S
%for the isotope $^{18}$C.}
%\label{1a}
%\end{figure}

\begin{figure}
\caption{Neutron particle (top) and pairing (bottom) densities of $^{18}$C
obtained with the zero-range pairing interaction, Eq. (\ref{12}), for
different values of the cut off.}
\label{2a}
\end{figure}

\begin{figure}
\caption{Upper part: 
  Total energies of $^{18}$C obtained with zero-range and Gogny 
pairing interactions for different cut off values. 
Lower part: Pairing correlation
energies, Eq. (\ref{13n}), obtained with zero-range (circles) and Gogny
(stars)  
pairing interactions for different cut
off values for $^{18}$C.}
\label{3a}
\end{figure}

%\begin{figure}
%\caption{Pairing correlation
%energies, Eq. (\ref{13n}), obtained with zero-range and Gogny 
%pairing interactions for different cutoff values for $^{18}$C.}
%\label{4a}
%\end{figure}

\begin{figure}
\caption{Mean square radii of pairing densities, 
  Eq. (\ref{14n}), obtained with zero-range (circles) and Gogny (stars)  
pairing interactions for different cut off values for $^{18}$C.}
\label{5a}
\end{figure}

\newpage

\includegraphics{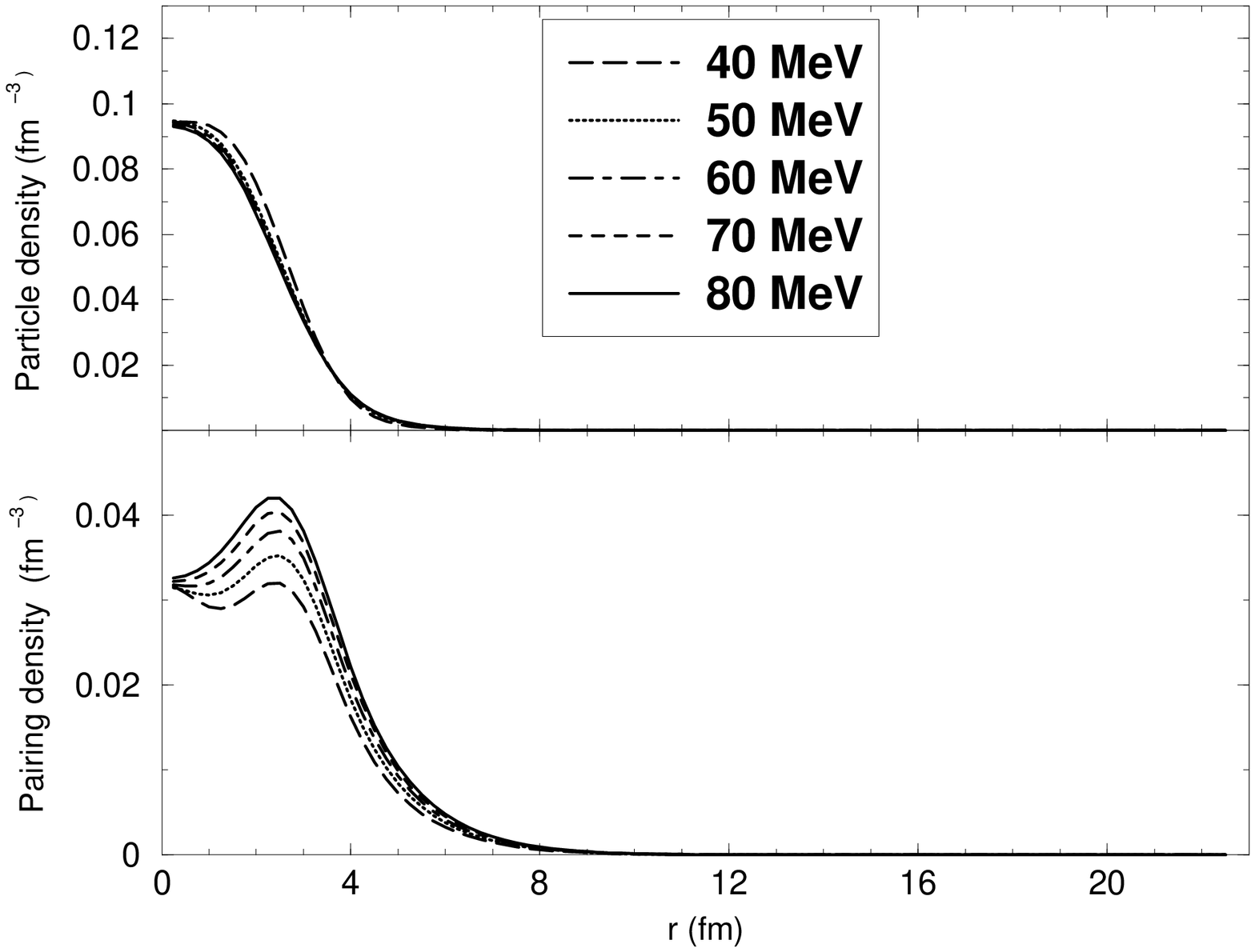}
\mbox{}\\

\newpage

\includegraphics{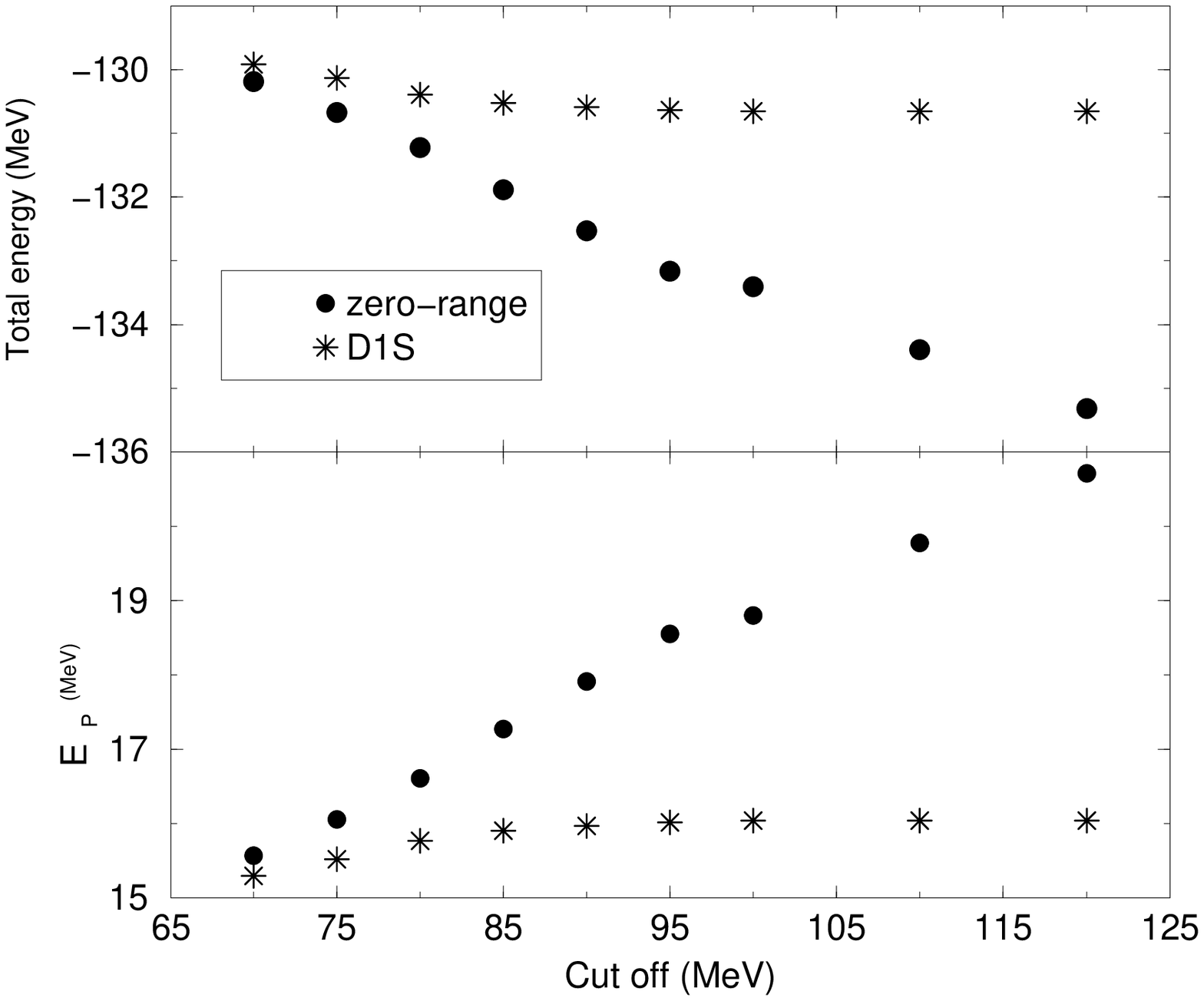}
\mbox{}\\

\newpage

\includegraphics{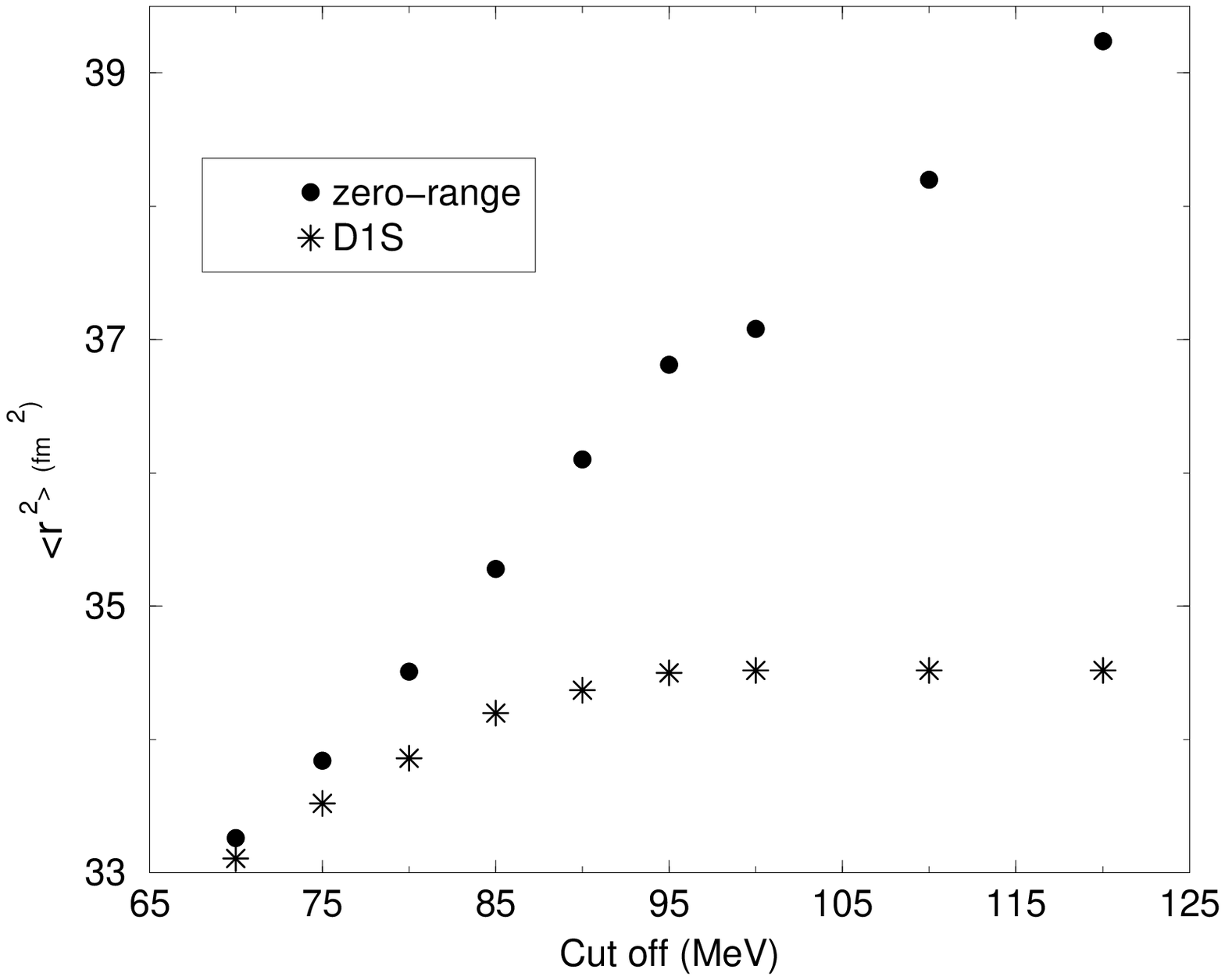}
\mbox{}\\

%\newpage

%\special{psfile=fig4.eps
%hoffset=-40 voffset=-500 hscale=90 vscale=80 angle=0}
%\mbox{}\\

%\newpage

%\special{psfile=figbis5.eps
%hoffset=-40 voffset=-500 hscale=90 vscale=80 angle=0}
%\mbox{}\\

%\newpage

\end{document}